\documentclass[twocolumn,showpacs,preprintnumbers,amsmath,amssymb]{revtex4}

\usepackage{graphicx,epsfig}
\usepackage{dcolumn}
\usepackage{bm}
\begin{document}


\title{A fresh look on the heating mechanisms of the Solar corona}

\author{David Tsiklauri}
\affiliation{Institute for Materials Research,
University of Salford, Gt Manchester, M5 4WT, United Kingdom}%
\date{\today}

\begin{abstract}

Recently using Particle-In-Cell simulations i.e. in the kinetic plasma description Tsiklauri et al. and G\'enot et al.  
reported on a discovery of a new mechanism of parallel electric field generation,
which results in electron acceleration. 
In this work we show that the parallel (to the uniform unperturbed magnetic field)
electric field generation can be obtained in much simpler framework
using ideal Magnetohydrodynamic (MHD) description, i.e. without resorting to complicated wave particle interaction effects  such as ion 
polarisation drift and resulting space charge separation which seems to be an ultimate cause of the electron acceleration. 
In the ideal MHD the parallel (to the uniform unperturbed magnetic field) electric field appears due to fast magnetosonic waves which are generated
by the interaction of weakly non-linear Alfv\'en waves with the transverse density inhomogeneity.
Further, in the context of the coronal heating problem  a new {\it two stage mechanism} of the plasma heating is
presented by putting emphasis, first, on the generation of parallel electric fields within {\it ideal MHD} description directly, 
rather than focusing on the enhanced dissipation mechanisms of the Alfv\'en waves and, second, dissipation of these parallel electric 
fields via {\it kinetic} effects. It is shown that a single 
 Alfv\'en wave harmonic  with frequency ($\nu = 7$ Hz), (which has longitudinal wavelength  $\lambda_A = 0.63$ Mm for putative
 Alfv\'en speed of 4328 km s$^{-1}$)
the generated parallel electric field could account for the 10\% of the necessary coronal heating requirement. 
We conjecture that  wide spectrum (10$^{-4}-10^3$ Hz) Alfv\'en waves, based on observationally constrained spectrum, could
provide necessary coronal heating requirement. It is also shown that the amplitude of generated parallel electric field 
exceeds the Dreicer electric field by about four orders of magnitude, which implies realisation of the run-away regime
with the associated electron acceleration.
\end{abstract}


\maketitle

During the total solar eclipse of August 7, 1869, Harkness and Young discovered an emission 
line of feeble intensity in the green part of the spectrum of the corona.
Similarly to the case of Helium discovered by Sir Norman Lockyer in 1868 
it was mistakenly proposed that this line was due to an unknown element, provisionally named coronium.
It was only in 1939 when Grotrian and Edlen correctly identified it as an emission
produced by highly ionised Iron at temperatures of a few $\times 10^6$ K.
Physical understanding of this high temperature in the solar corona is still
a fundamental problem in astrophysics. None of the existing models
can simultaneously account for all the observational and physical requirements
in order to explain what is known as the coronal heating problem.

The second law of thermodynamics states that: 
It is not possible for heat to flow from a colder body to a warmer body without any 
work having been done to accomplish this flow. Or
energy will not flow spontaneously from a low temperature object to a higher temperature object. 
This seems to be in apparent contradiction with what we see in the solar atmosphere where
temperature steeply rises from the photospheric boundary which is at $T=5785$ K
to a few $\times 10^6$ K i.e. an increase of a factor of 200.
The transition region between the 
chromosphere (a thin layer above photosphere) and the hot corona 
over which the temperature increases dramatically
is extremely thin ($\approx$ 0.01 \% of the Sun's diameter).
This implies that some intense heat deposition occurs in the corona such that the second law of
thermodynamics can be saved.

The temperature structure of the solar corona is far from homogeneous \cite{a04}.
The optically thin emission from the corona in soft X-rays or in the extreme ultra violet
implies over-dense structures: so called coronal loops (closed magnetic structures) or 
plumes in polar regions (open magnetic structures) amongst others.
In the corona the thermal pressure is generally much smaller than 
the magnetic pressure so that their ratio which is known as plasma-$\beta$ parameter
is $\ll 1$. This implies that in the corona 
substantial amount of energy is stored in the magnetic field.
In turn, if one could devise a mechanism how this energy is released (dissipated),
then important clues to the solution of the coronal heating 
problem would be revealed.

The coronal heating models are subdivided according to the  possibilities of
how the currents that are responsible for the plasma heating are dissipated: 
either by magnetic reconnection \cite{p83,p03,plt03}, 
Ohmic dissipation via current cascading \cite{vb86,gn96}, and viscous 
turbulence \cite{hp92} in the case of so called DC models \cite{a04}, or 
by Alfv\'enic resonance, i.e. resonant
absorption \cite{i78,ods95,eg96,vh04}, 
phase mixing \cite{hp83,nrm97,rnr98,bank00,tan01,hbw02,tn02,tna02,tnr03,tss05a,tss05b}, 
and turbulence \cite{ip95} in the case of AC models.
An interesting alternative to all of the above models was explored  in \cite{t05},
where coronal loops could be 
heated by the flow of solar wind plasma (plus other flows that may be present) across them  by generating
currents in a similar manner as a conventional Magnetohydrodynamic (MHD) generator. 

Historically all phase mixing studies have been performed in the Magnetohydrodynamic (MHD) approximation,
however, since the transverse scales in the Alfv\'en wave collapse progressively to zero,
the MHD approximation is inevitably violated. 
Thus, Refs.\cite{tss05a,tss05b} studied the phase mixing effect in the kinetic regime, i.e.
beyond a MHD approximation, where
a new mechanism for the acceleration of electrons
due to the generation of parallel electric field in the solar coronal context was discovered. 
This mechanism has important implications
for various space and laboratory plasmas, e.g. the 
coronal heating problem and acceleration of the solar wind.
It turns out that in the magnetospheric context similar parallel electric field generation 
mechanism in the transversely inhomogeneous plasmas was reported before
\cite{glm04,glq99} (see also the comment letter \cite{mgl06} and references therein).

Contrary to the previous studies \cite{tss05a,tss05b,glm04,glq99}, here we use MHD description of the problem.
Namely we solve numerically ideal, 2.5D, MHD equations in Cartesian coordinates, 
with plasma $\beta=0.0001$ starting from the following equilibrium
configuration: A uniform magnetic field $B_0$ in the $z-$direction penetrates plasma with the density
inhomogeneity across $x-$direction, which is varying such that it
increases from 
some reference background value of $\rho_0$,
which in our case was fixed at 
$\rho_0=2\times10^9 \mu m_p$ g cm$^{-3}$ (with molecular weight of 
$\mu=1.27$ corresponding to the solar coronal conditions $^1$H:$^4$He=10:1 \cite{a04}
and $m_p$ being the proton mass), to $5 \rho_0$.
Such density profile across the magnetic field has steep gradients with half-width of 3 Mm around
$x \simeq \pm 10$ Mm and is essentially flat elsewhere. Here "M" in units stands for mega i.e. $10^6$. Such a structure mimics e.g.
footpoint of a large curvature radius solar coronal loop or a polar
region plume  with the ratio of density inhomogeneity scale and the loop/plume radius of 0.3, which is
a median value of the observed range 0.15 - 0.5 \citep{g2002}.  We use the following usual normalisation 
$B_{x,y,z}=B_0 \bar{B}_{x,y,z}$,
$(x,y,z)=a_0(\bar{x},\bar{y},\bar{z})$, $t=(a_0/c_A^0)\bar{t}$.
Here $B_0$ we fix at 100 G, and hence dimensional Alfv\'en speed, 
$c_A^0=B_0/\sqrt{4 \pi \rho_0}$
turns out to be 4328 km s$^{-1}$=0.0144 $c$ ($c$ is the speed of light). 
The reference length $a_0$ was fixed at 
1 Mm, i.e. dimensionless time of unity corresponds to  0.2311 s.
We usually omit bar on top of normalised quantities, hence when
simply numbers are quoted without units, in such circumstances
we refer to dimensionless units as defined above.
The dimensionless Alfv\'en speed is normalised to $c_A^0$.
The simulation domain spans from $-40$Mm to 40 Mm in both $x-$ and $z-$directions
with the density ramp as described above mimicking a footpoint fragment
of a solar coronal loop  or a polar region plume. Our initial equilibrium is depicted in Fig.~1.

The initial conditions for the numerical simulation are
$B_y=A \cos(kz)$ and $V_y=-c_A(x)B_y$ at $t=0$, which means that purely 
Alfv\'enic, linearly polarised, plane wave is launched travelling in the
direction of positive $z$'s. The rest of physical quantities, namely,
 $V_x$ and $B_x$ (which 
would be components of fast magnetosonic wave if the medium were totally homogeneous)
and $V_z$ and $B_z$ (the analogs of slow magnetosonic wave) are initially set
to zero. 
Plasma temperature is varied as inverse of density so that the total (thermal plus magnetic) pressure always
remains constant.
We fixed the amplitude of  Alfv\'en wave $A$ at 0.05. This choice makes
Alfv\'en wave weakly non-linear. 

The models of coronal heating using wave dissipation were focusing
on the mechanisms (e.g. phase mixing) 
that could enhance damping of the Alfv\'en waves.
However for the coronal value of shear (Braginskii) viscosity,
by which Alfv\'en waves damp of about
$\eta = 1$ m$^2$ s$^{-1}$, typical dissipation lengths ($e$-fold decrease of Alfv\'en wave amplitude
over that lengths) are $\simeq 1000$ Mm, 
only invoking somewhat ad hoc concept of so called enhanced resistivity
can bring down the dissipation length to a required value of the 
order of hydrodynamic pressure scale height $\lambda_T \approx 50$ Mm.
In this light (observation of mostly undamped Alfv\'en waves \cite{moran01} (see however Ref.\cite{obj05})
and inability of classical (Braginskii) viscosity producing short enough
dissipation length), it seems reasonable to focus rather on the generation
of parallel electric fields which would guarantee plasma heating should
the energy density of the parallel electric fields be large enough.
We put emphasis on parallel electric fields because in the direction parallel 
to the magnetic field electrons (and protons) are not constrained.
While in the perpendicular to the magnetic field direction
particles are constrained because of large classical conductivity 
$\sigma=6 \times 10^{16}$ s$^{-1}$ (for $T=2 MK$ corona) and inhibited 
momentum transport across the magnetic field.

In the MHD limit there are no parallel electric fields associated with the 
Alfv\'en wave. In the single fluid, ideal MHD 
(which is justified because of large $\sigma$) the parallel (to the uniform unperturbed magnetic field)
electric field 
can be obtained from 
\begin{equation}
E_z = - \frac{V_x B_y - V_y B_x}{c}.
\end{equation}
This means that in the considered system 
$E_z$ can only be generated if the initial
Alfv\'en wave ($V_y,B_y$) is able to generate 
fast magnetosonic wave ($V_x,B_x$).

In the recent past, in the context other than parallel electric field generation,
Ref. \cite{nrm97} investigated just such 
a possibility of growth
of fast magnetosonic waves in a similar physical system. They
used mostly analytical 
approach and focused on the early stages of the system's evolution.
Later, long term evolution of the fast magnetosonic wave
 generation was studied numerically  in the case
of harmonic \cite{bank00} and Gaussian \cite{tan01} Alfv\'enic
initial perturbations. 
When fast
magnetoacoustic perturbations are initially absent 
and the initial
amplitude of the plane Alfv\'en wave is small, 
the subsequent evolution
of the wave, due to the difference in local Alfv\'en speed
across the $x$-coordinate, leads to the distortion of the wave
front. This leads to the appearance of transverse (with respect to
the applied magnetic field) gradients, which grow linearly
with time. 
The main negative outcome of the studies with the 
harmonic \cite{bank00} and Gaussian \cite{tan01} Alfv\'enic
initial perturbations
was that the amplitudes of $V_x$ and $B_x$
 after rapid initial growth actually tend to
saturate due to the destructive wave interference effect \cite{tan01}.

We used fully non-linear 
MHD code Lare2d with the initial conditions as described above.
As a self-consistency test  we performed numerical simulations with moderate values of 
$k=1$ (in dimensional units this corresponds to an 
Alfv\'en wave with frequency ($\nu = 0.7$ Hz), i.e. longitudinal wave-numbers $\lambda_A = 6.3$ Mm) and corroborated previous results of 
\cite{bank00,tan01}.

In Fig.~(2) we show two snapshots of $V_x$ and $E_z$ (the latter was reconstructed using Eq.(1))
for the case of $k=10$ (in dimensional units this corresponds to an 
Alfv\'en wave with frequency ($\nu = 7$ Hz), i.e. longitudinal wave-number $\lambda_A = 0.63$ Mm.).
It can be seen that the fast magnetosonic wave ($V_x$) and parallel electric
field ($E_z$) are both generated in the vicinity of the density gradients $x \simeq \pm10$,
eventually filling the entire density ramp. This means the generated parallel electric
fields are confined by the density gradients, i.e. the solar coronal loop (which the considered
systems tries to mimic) after about 20 Alfv\'en time scales becomes filled with the
oscillatory in time parallel electric fields. 
Also, as in 
\cite{tss05b} and \cite{glm04}, the generated parallel electric field
is quite spiky, but more importantly it seems that 
large wave-numbers (and frequencies) i.e. short wavelength now are capable to
significantly increase the amplitudes of generated
both fast magnetosonic waves ($V_x$) and parallel electric field $E_z$ shown in Fig.~(3).
In Fig.~(3) we plot the amplitudes of $V_x \equiv V_x^a$ and $E_z \equiv E_z^a$ which we 
define as the  maxima of absolute values of the wave amplitudes along $x \simeq \pm10$ line
(which essentially track the generated wave amplitudes in the strongest density gradient regions) 
at a given time step. This amplitude 
growth is beyond simple $A^2$ scaling (cf. \cite{tan01}).
In the considered case $E_z$ now attains values of 0.001.
To verify the convergence of the solution, we plot the results of the
numerical run with the doubled ($4000 \times 4000$) spatial resolution.
The match seems satisfactory, which validates the obtained results.
One could argue that at first sight, the  obtained spikes in 
both fast magnetosonic waves ($V_x$) and parallel electric field $E_z$
do not contain much energy as they are localised strongly in the wavefront.
However, one should realise that in Fig.~(2) case of a single harmonic with 7 Hz is considered.
If one considers a {\it wide spectrum} of Alfv\'en waves (see discussion of this conjecture below), 
then the harmonics with different frequencies will contribute to the generation of what will then be a "forest"
of such spikes (for each harmonic). Also, as the frequency of Alfv\'en waves decreases than much more regular (no longer spiky)
wave structures are observed for both fast magnetosonic waves ($V_x$) and parallel electric field $E_z$.
To demonstrate this point we present intensity plots of $V_x$ and $E_z$ at 20 for the case of $k=1$, $\nu = 0.7$ Hz, $\lambda_A = 6.3$ Mm
in Fig.~4. In the latter case the amplitude of $V_x$ saturates at $A^2=0.05^2=0.0025$ as one might expect from the weakly
non-linear theory, while $E_z$ at $3\times 10^{-5}$.

It has been know for decades \cite{kis81} that the coronal energy losses that need to be compensated
by some additional energy input, to keep solar corona to the observed temperatures, are as following (in units of erg cm$^{-2}$ s$^{-1}$):
$3\times10^5$ for quiet Sun, $8\times10^5$ for a coronal hole and $10^7$ for an active region.
Ref.\cite{a04} makes similar estimates for the heating flux per unit area (i.e. in erg cm$^{-2}$ s$^{-1}$):
\begin{equation}
F_H=E_H \lambda_T=5 \times 10^3 \left(\frac{n_e}{10^8 {\rm cm}}\right)^2\left(\frac{T}{1 {\rm MK}}\right),
\end{equation}
where $E_H \approx 10^{-6}$ erg cm$^{-3}$ s$^{-1}$. This yields an estimate of $F_H \approx 2\times10^6$ erg cm$^{-2}$ s$^{-1}$
in an active region of the corona with a typical loop base electron number density of $n_e=2\times10^9$ cm$^{-3}$ and $T=1$ MK.

In order to make appropriate estimates we first note that the energy density associated
with the parallel electric field $E_z$ is $E_E= \varepsilon E_z^2/(8 \pi)$
 erg cm$^{-3}$,
where $\varepsilon$ is the dielectric permittivity of plasma. The latter can be calculated from 
$\varepsilon =(4 \pi \rho c^2)/B^2$.
This expression for $\varepsilon$ is different from the usual expression for the 
dielectric permittivity \cite{kt73}: $\varepsilon=1+(4 \pi \rho c^2)/B^2$
because the displacement current has been neglected in the above treatment, which is a usual
assumption made in the MHD limit. At any rate for the coronal conditions ($\rho=2\times10^9 \mu m_p$ g cm$^{-3}$, 
$\mu=1.27$, $B=100$ Gauss) the second term $(4 \pi \rho c^2)/B^2=4.8048 \times 10^3 \approx \varepsilon \gg 1$.
>From Fig.~(3) we gather that the parallel electric field amplitude attains value of $\approx 0.001$. In order to convert
this to dimensional units we use $c_A^0=4328$ km s$^{-1}$ and $B=100$ G and Eq.(1) to obtain
$E_z\approx(c_A^0 B/c)\times 0.001=0.0014$ statvolt cm$^{-1}$ (in Gaussian units).
Therefore the energy density associated
with the parallel electric field $E_z$  is $E_E=(4.8048 \times 10^3) \times 0.0014^2/(8 \pi)=3.7471 \times 10^{-4}$
erg cm$^{-3}$.
In order to get  the heating flux per unit area for a {\it single harmonic} with frequency
7 Hz, we multiply the latter expression by the 
Alfv\'en speed of 4328 km s$^{-1}$ (this is natural step because the fast magnetosonic waves ($V_x$ and $B_x$) which propagate across
the magnetic field and associated parallel electric fields ($E_z$) are generated on density gradients by the Alfv\'en waves. 
Hence, the flux is carried with the {\it Alfv\'en} speed)  to obtain
\begin{equation}
F_E=E_E c_A^0=1.62\times10^5\;\;\;
\left[{\rm erg \, cm^{-2} s^{-1}}\right],
\end{equation}
which is $\approx 10$ \% of the
coronal heating requirement estimate for the same parameters made above using Eq.(2).
Note that the latter estimate is for  a {\it single harmonic} with frequency
7 Hz (see discussion below for details when a wide spectrum of Alfv\'en waves is considered).

We now discuss how the energy stored in the generated parallel electric field is dissipated. 
For this purpose the parallel electric field behaviour at a given point in space (as a function of
time) was studied. We found that $E_z$  (in the point of strongest density gradient) is periodic (sign-changing) function which is a mixture
of two harmonics with frequencies $\omega=c_A k$ and $\omega=2 c_A k$. This can be qualitatively explained
by the fact that $E_z$ is calculated using Eq.(1) where $V_y$ and $B_y$ at fixed spatial point
vary in time with frequencies $\omega=c_A k$, while the generated $V_x$ and $B_x$
do so with frequencies $\omega=2 c_A k$ \cite{nrm97}.
Under influence of such periodic parallel electric field electrons start
oscillations and are accelerated (while ions perhaps are not due to their larger inertia).
It should be emphasised that because of the ideal MHD approximation used in this letter, strictly speaking
the generated electric field cannot accelerate plasma particles or cause Ohmic heating {\it unless kinetic effects are
invoked}. Let us look at the Ohm's law for ideal MHD (Eq.(1)) in more detail denoting unperturbed by the waves
physical quantities with subscript 0 and the ones associated with the waves by a prime:
Initial equilibrium implies $\vec E_0= \vec V_0 =0$ with $\vec B_0 \not = 0$. For the perturbed state 
(with Alfv\'en waves ($V_y$ and $B_y$) launched which generate fast magnetosonic waves ($V_x$ and $B_x$) ) 
we have 
\begin{equation}
\vec E^{\prime}=- {\vec V^{\prime} \times (\vec B_0 +\vec B^{\prime})}/{c}
\end{equation}      
Note that the projection of $\vec E^\prime$ on full magnetic field (unperturbed $\vec B_0$ plus the waves $\vec B^\prime$)
is zero by the definition of cross and scalar product: 
$\vec E^ \prime \cdot (\vec B_0 +\vec B^\prime)/ |(\vec B_0 +\vec B^\prime)| =0$. Physically this means that
in ideal MHD electric field cannot do any work as it is always perpendicular to the {\it full} (background plus wave) magnetic field.
However the projection of $\vec E^\prime$ on unperturbed magnetic field $\vec B_0$ is clearly non-zero
\begin{equation}
\vec E^ \prime \cdot \frac{\vec B_0}{ |\vec B_0 |}= E_z=-\frac{\vec V^{\prime} \times \vec B^{\prime} }{c}
\cdot \frac{\vec B_0}{ |\vec B_0 |}
\end{equation}
which exactly coincides with Eq.(1) that was used to calculate $E_z$ throughout this letter.
Crucial next step needed to understand how the generated parallel (to the uniform unperturbed magnetic field)
electric fields dissipate {\it must invoke kinetic effects}. In our two stage model, in the first stage bulk MHD motions (waves) are
generating the parallel electric fields, which as we saw cannot accelerate particles if we describe plasmas in the ideal MHD
limit. Clearly we witnessed in works of \citet{glm04} and \citet{tss05b} that when identical system is
modelled in kinetic regime particles are indeed accelerated with such parallel fields.
\citet{glm04} claimed that electron acceleration is due to the polarisation drift.
In particular, they showed that once Alfv\'en wave propagates on the
transverse to the magnetic field density gradient, parallel electric field is generated due to
charge separation caused by the polarisation drift associated with the
time varying electric field of the Alfv\'en wave. Because polarisation drift speed is
proportional to the mass of the particle, it is negligible for electrons, hence ions
start to drift. This causes charge separation (the effect that is certainly beyond reach of MHD
description), which results in generation of the parallel electric fields, that in turn accelerates electrons.
In the MHD consideration our parallel electric field is also time varying, hence at the second (kinetic) stage the electron
acceleration can proceed in the same manner through the ion polarisation drift and charge separation.
The exact picture of the particle dynamics in this case can no longer be treated with MHD and kinetic
description is more relevant. 
It should be mentioned that the frequencies considered in kinetic regime \citep{tss05b}, $\nu=0.3 \omega_{ci}/(2 \pi)=4.6\times10^4$ Hz
and this letter, $0.7-7$ Hz, are different, but we clearly saw that the increase in frequency just yields enhancement of the
parallel electric field generation.
Various effects of wave-particle interactions will rapidly damp the
parallel electric fields on a time scale much shorter than MHD time scale.
\citet{glm04} clearly demonstrated the role of nonlinearity and kinetic instabilities in the
rapid conversion from initial low frequency electromagnetic regime, to a high frequency
electrostatic one. They identified Buneman and weak beam plasma instabilities in their simulations
(as they studied time evolution of the system for longer than \citet{tss05b} times).
Fig.(11) from \citet{glm04} shows that wave energy is converted into particle energy
on the times scales of $10^3 \omega_{pe}^{-1}\approx 4$ Alfv\'en periods.
Perhaps no immediate comparison is possible (because of the different frequencies involved), but still in our case Alfv\'en period when sufficient energy is
stored in the parallel electric field is $1/(7$ Hz)=0.14 s, i.e. wave energy is converted into the particle energy in quite short time.

Our aim here was to show that parallel electric field {\it generation } is
possible even within the MHD framework without resorting to plasma kinetics.
 Yet another important observation can be made by estimating the Dreicer electric field \cite{dreicer}.
Dreicer considered dynamics of electrons under action of two effects: the parallel electric field and friction between electrons and ions.
He noted that the equation describing electron dynamics along the magnetic field can be written as
\begin{equation}
m_e \dot{v_d}=eE-\nu_p^{e/i}m_ev_d
\end{equation}
where $v_d$ is the electron drift velocity and $\nu_p^{e/i}$ is the electron collision frequency and dot denotes time derivative.
We note in passing that when $v_d \ll v_{thermal}$ Eq.(6) in the steady state regime ($d/dt=0$) allows to 
derive the expression for Spitzer resistivity. When $v_d > v_{thermal}$ then steady state solution may not apply.
In this case when the right hand side of Eq.(6) is positive i.e. when $eE>\nu_p^{e/i}m_ev_d$,
we have electron acceleration. In fact this is so called {\it run-away regime}.
In simple terms acceleration due to parallel electric field leads to an increase in $v_d$, in turn, this leads to decrease in $\nu_p^{e/i}$
because $\nu_p^{e/i}\propto 1/v_d^3$. In other words when electric field exceeds the critical value $E_d=n_ee^3\ln\Lambda/(8 \pi \varepsilon_0^2k_BT)$ 
(the Dreicer electric field, which is quoted here in SI form), 
faster drift leads to decrease in electron-ion friction, which in turn results in even faster drift speed and
hence  run-away regime is reached. Putting in coronal values ($n_e=2\times10^9$ cm$^{-3}$, $T=1$ MK and $\ln\Lambda=17.75$) 
in the expression for the Dreicer electric field we obtain 0.0054 V m$^{-1}$
which in Gaussian units is $1.8\times 10^{-7}$ statvolt cm$^{-1}$.
As can be seen from this estimate the values of parallel electric field obtained in this letter exceeds 
the Dreicer electric field by about four orders of magnitude! This guarantees that ran-away regime would take place
leading to the electron acceleration and fast conversion of the generated electric field energy in to heat.
It should be noted, however, that in Eq.(6) $E$ is constant, while our $E_z$ in the point of strongest density gradient is
time varying. Hence some modification from the Dreicer analysis is expected. 

It has been known for decades that the flux carried by Alfv\'en waves in
the solar corona is more than enough to heat it to the observed temperatures \cite{kis81}, however
linear MHD  Alfv\'en waves in homogeneous plasma do not possess parallel electric field component. This is one of the reasons 
why they are so difficult to dissipate. Hence the novelty of this study was 
to consider situation when parallel electric field generation is possible (weak non-linearity and transverse density inhomogeneity).

After comment paper by \citet{mgl06} we came to realisation that electron acceleration seen in both series of 
works \citep{tss05a,tss05b,glm04,glq99} is a non-resonant wave-particle interaction effect. In works by \citet{tss05a,tss05b}
electron thermal speed was $v_{th,e}=0.1c$ while Alfv\'en speed in the strongest density gradient regions
was $v_A=0.16c$, this unfortunate coincidence led us then to the conclusion that the electron acceleration by parallel
electric fields was affected by the Landau resonance with the phase mixed Alfv\'en wave. In works by \citep{glm04,glq99}
electron thermal speed was $v_{th,e}=0.1c$ while Alfv\'en speed was $v_A=0.4c$ because they considered more
strongly magnetised plasma applicable to Earth magnetospheric conditions. 
There were three main stages that lead to the formulation of the present model.

(i) The realisation of the parallel electric field generation (and particle acceleration) being
 a {\it non-resonant} wave-particle interaction effect, lead us to a question: 
 could such  parallel electric fields be generated in MHD approximation? 

(ii) Next we realised that indeed if one considers
{\it non-linear} generation of fast magnetosonic waves in the plasma with 
transverse density inhomogeneity, then $\vec E = - (\vec V \times \vec B)/c$ contains non-zero component
parallel to the ambient magnetic field $E_z = - (V_x B_y - V_y B_x)/c$.  

(iii) From previous studies \citep{bank00,tan01} we knew that the fast magnetosonic waves ($V_x$ and $B_x$) did not grow to
a substantial fraction of the Alfv\'en wave amplitude. However after reproducing old parameter regime (k=1, i.e frequency of 0.7 Hz),
fortunately case of k=10, i.e frequency of 7 Hz was considered, which showed that  fast magnetosonic waves
and in turn parallel electric field were more efficiently generated.

We close this letter with the following four points:

(i) In this work we showed that the parallel electric field generation reported in number of previous
publications \cite{tss05a,tss05b,glm04,glq99} which dealt with transversely inhomogeneous plasma can be explained in much simpler framework
using MHD description, i.e. without resorting to complicated wave particle interaction effects.

(ii)
In the context of the coronal heating problem  a new {\it two stage mechanism} of the plasma heating is
presented by putting emphasis, first, on the generation of parallel electric fields within {\it ideal MHD} description directly, 
rather than focusing on the enhanced dissipation mechanisms of the Alfv\'en waves and, second, dissipation of these parallel electric 
fields via {\it kinetic} effects. 

(iii) 
It is shown that a single 
 Alfv\'en wave harmonic  with frequency ($\nu = 7$ Hz), (which has longitudinal wavelength  $\lambda_A = 0.63$ Mm for putative
 Alfv\'en speed of 4328 km s$^{-1}$)
the generated parallel electric field could account for the 10\% of the necessary coronal heating requirement. 
We conjecture that the wide spectrum (10$^{-4}-10^3$ Hz) Alfv\'en waves, based on observationally constrained spectrum, could
provide necessary coronal heating requirement. 
In this regard, it should mentioned that 
Alfv\'en waves as observed in situ in the solar wind always appear to be
propagating away from the Sun and it is therefore natural to assume a solar origin for
these fluctuations. However, the precise origin in the solar atmosphere of the hypothetical
source spectrum for Alfv\'enic waves (turbulence) is unknown, given the impossibility of remote
magnetic field observations above the chromosphere-corona transition region \citep{vp97}.
Studies of ion cyclotron resonance heating of the solar corona and high speed winds exist which
provide important spectroscopic constraints on the Alfv\'en wave spectrum \citep{cfk99}.
Although the spectrum can and is observed at distances of 0.3 AU, it can be then projected back at the base of
corona using empirical constraints, see e.g. top line in Fig. 5 from \citet{cfk99}.
Using the latter figure we can make the following estimates. Let us look at single harmonic, first. At frequency
7 Hz (used in our simulations) magnetic energy of Alfv\'enic fluctuations is $E_\nu^{(\rm 7 \, Hz)} \approx 10^7$ nT$^2$ Hz$^{-1}$. For single harmonic 
with $\nu=7$ Hz this gives for the energy density $E^{(\rm 7 \, Hz)} \equiv \nu E_\nu^{(\rm 7 \, Hz)}/ (8 \pi) \approx 7 \times 10^{-3}$ G$^2/(8 \pi)\approx 2.8 \times 10^{-4}$ 
erg cm$^{-3}$. Note that surprisingly this {\it semi-observational} value  is quite close to our {\it theoretical} value of
$3.7471 \times 10^{-4}$ erg cm$^{-3}$!
As we saw above such single harmonic can provide approximately 10 \% of the coronal heating requirement.
Next let us look at how much energy density is stored in Alfv\'en wave spectrum based on empirically guided 
top line in Fig. 5 from \citet{cfk99}. 
Their spectral energy density (which they call "power") is approximated by so called $1/f$ spectrum, i.e.
$E_\nu=0.6 \times 10^8 / \nu$ nT$^2$ Hz$^{-1}$. Which in proper energy density units is $E_\nu=2.4 \times 10 ^{-4} / \nu$
erg cm$^{-3}$ Hz $^{-1}$. Hence the flux carried by  Alfv\'en waves from say $10^{-4}$ Hz {\it up to} a frequency $\nu$
would be 
\begin{equation}
F^{\rm AW}= \int_{10^{-4} \, {\rm Hz}}^{\nu}E_\nu c_A^0 d \nu= 1.04 \times 10^5 \ln(\nu/10 ^{-4})
\end{equation}
$\left[{\rm erg \, cm^{-2} s^{-1}}\right]$.
Based on the latter equation we deduce that the Alfv\'en wave spectrum from 
$\nu=10^{-4}$ Hz up to about
few $\times 10^3$ Hz carries a flux that is nearly as much as the coronal heating requirement, $F_H \approx 2\times10^6$ erg cm$^{-2}$ s$^{-1}$, quoted above.
We refrain from considering higher frequencies because for about $10^4$ Hz ions become resonant with circularly
polarised  Alfv\'en waves and the dissipation proceeds through the Landau resonance -- the well studied mechanism, but quite 
different from our non-resonant mechanism of parallel electric field generation.

There are several possibilities how this flux carried by Alfv\'en waves (fluctuations), 
is dissipated. If we consider regime of frequencies up to $10^3$ Hz, ion cyclotron resonance condition is not
met and hence dissipation would be dominated through the mechanism of parallel electric field dissipation formulated in this
letter. However at this stage it is unclear how much energy could actually be dissipated. This is due to the fact that we
only have two points 0.7 Hz and 7 Hz in our "theoretical spectrum". As we saw a single Alfv\'en wave harmonic with frequency 
7 Hz can dissipate enough heat to account for 10\% of the coronal heating requirement. 
Therefore, we conjecture that  wide spectrum (10$^{-4}-10^3$ Hz) Alfv\'en waves, based on observationally constrained spectrum, could
provide necessary coronal heating requirement. More details on the issue of wide spectrum will be published elsewhere \cite{t06}.

(iv) The obtained value of the generated parallel electric field 
exceeds the Dreicer electric field by about four orders of magnitude, which implies realisation of the run-away regime
with the associated electron acceleration.

{\it acknowledgements
Author kindly acknowledges support from the Nuffield Foundation (UK) through an award to newly 
appointed lecturers in Science, Engineering and Mathematics (NUF-NAL 04) and from the 
University of Salford Research Investment Fund 2005 grant.
Author acknowledges use of E. Copson Math cluster funded by PPARC 
and University of St. Andrews. Author also would like to thank two anonymous referees for the comments that improved this letter.}


\newpage
\begin{figure}[]
\resizebox{\hsize}{!}{\includegraphics{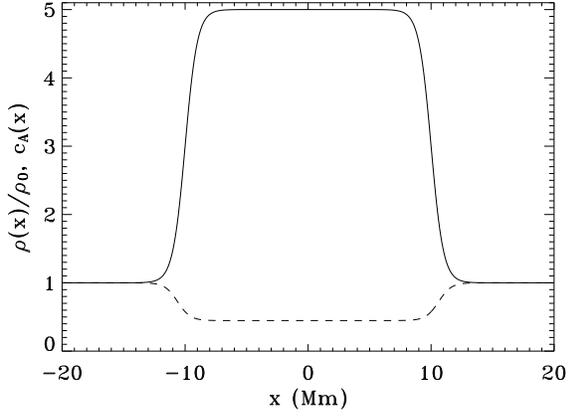}} 
\caption{Dimensionless density ($\rho(x)/ \rho_0 = \left[1+2\left(\tanh(x+10)+\tanh(-x+10)\right)\right]$) (solid line) and Alfv\'en speed, 
($c_A(x)=1 / \sqrt{1+2\left(\tanh(x+10)+\tanh(-x+10)\right)}$) (dashed line) profiles across the uniform unperturbed magnetic field (i.e. along $x$-coordinate) which is
used as an equilibrium configuration in our model of a footpoint of a solar coronal loop or a solar polar region plume.}
\end{figure}

\begin{figure*}
\centering
\epsfig{file=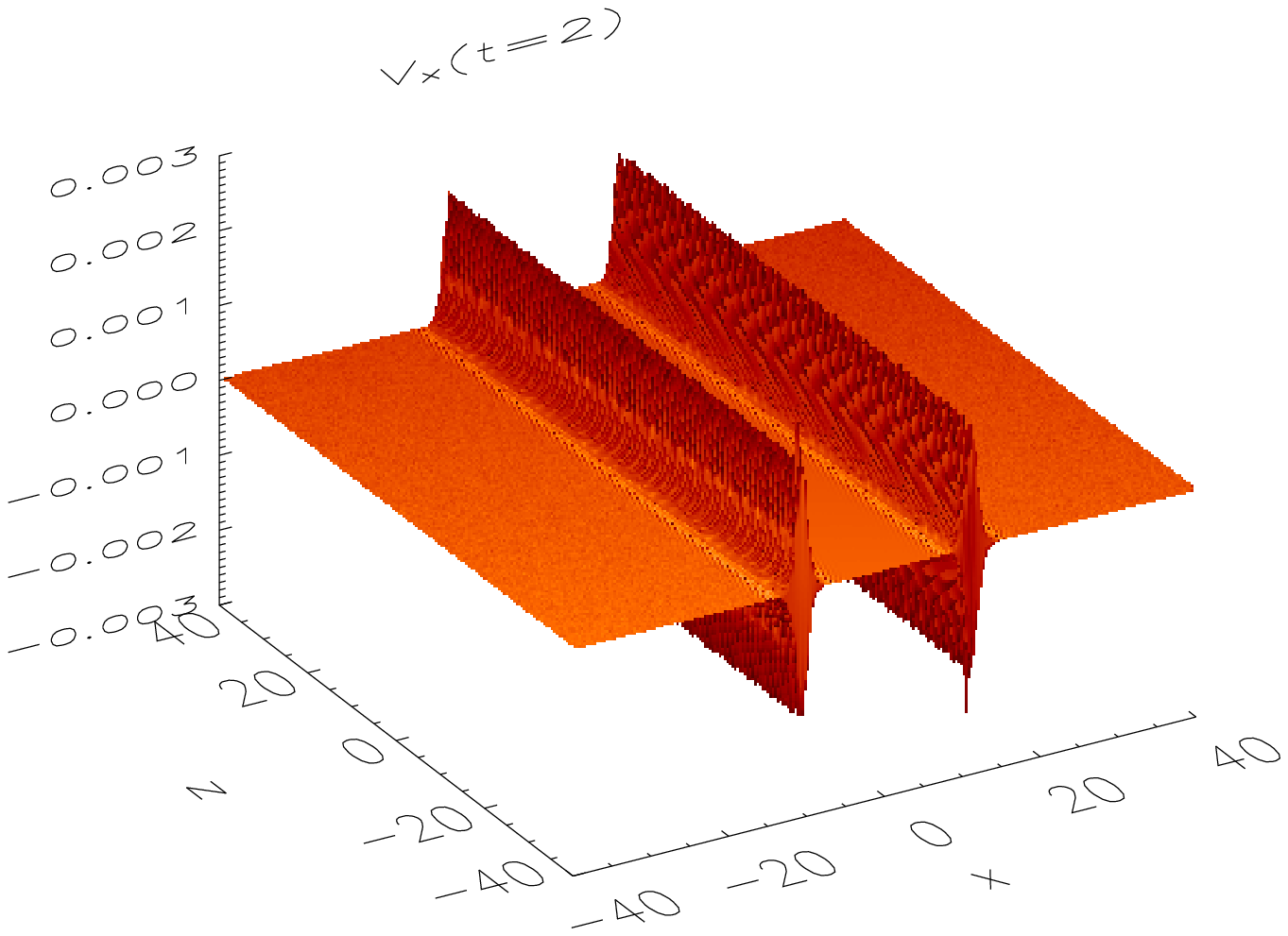,width=6.5cm}
 \epsfig{file=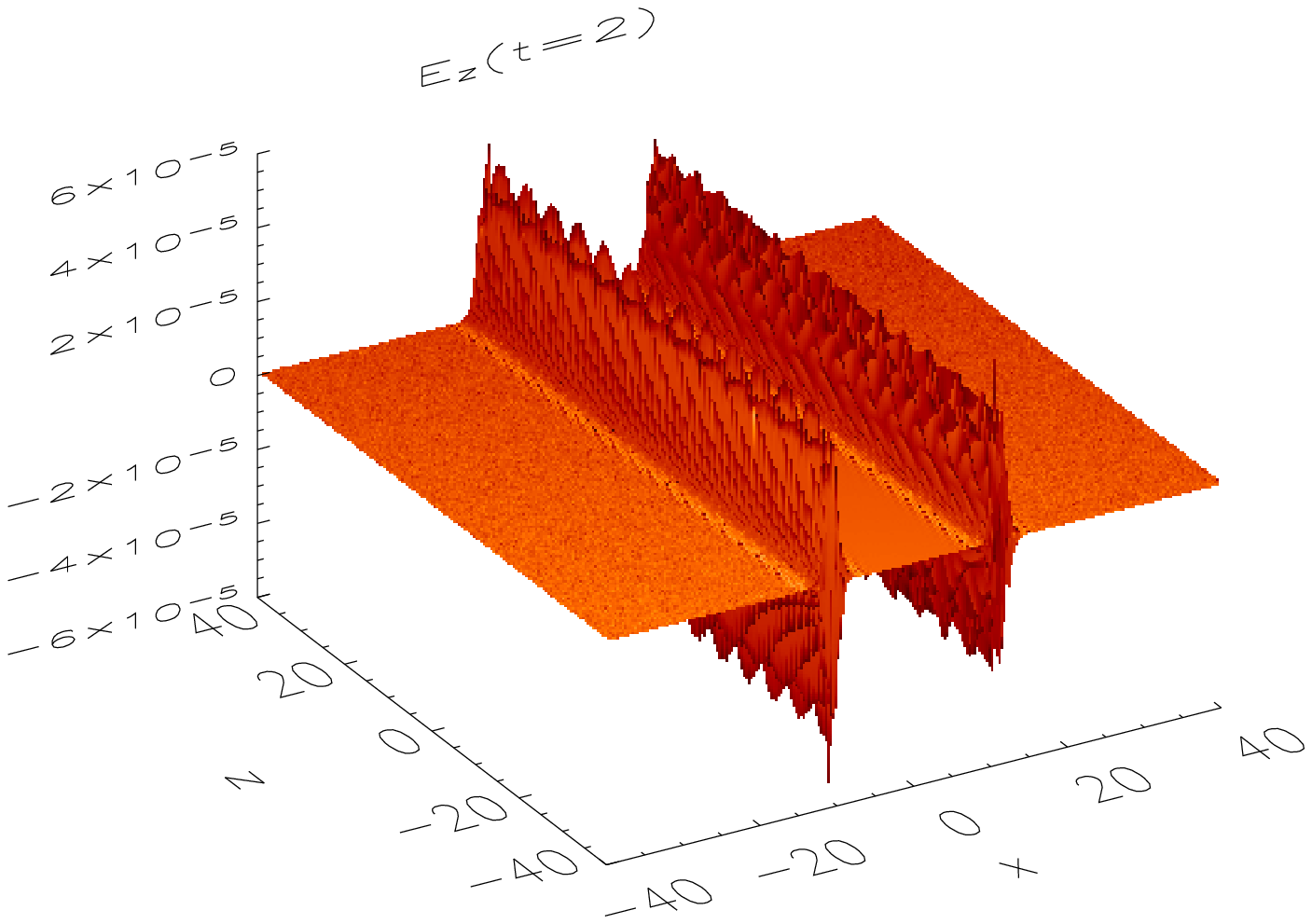,width=6.5cm}
 \epsfig{file=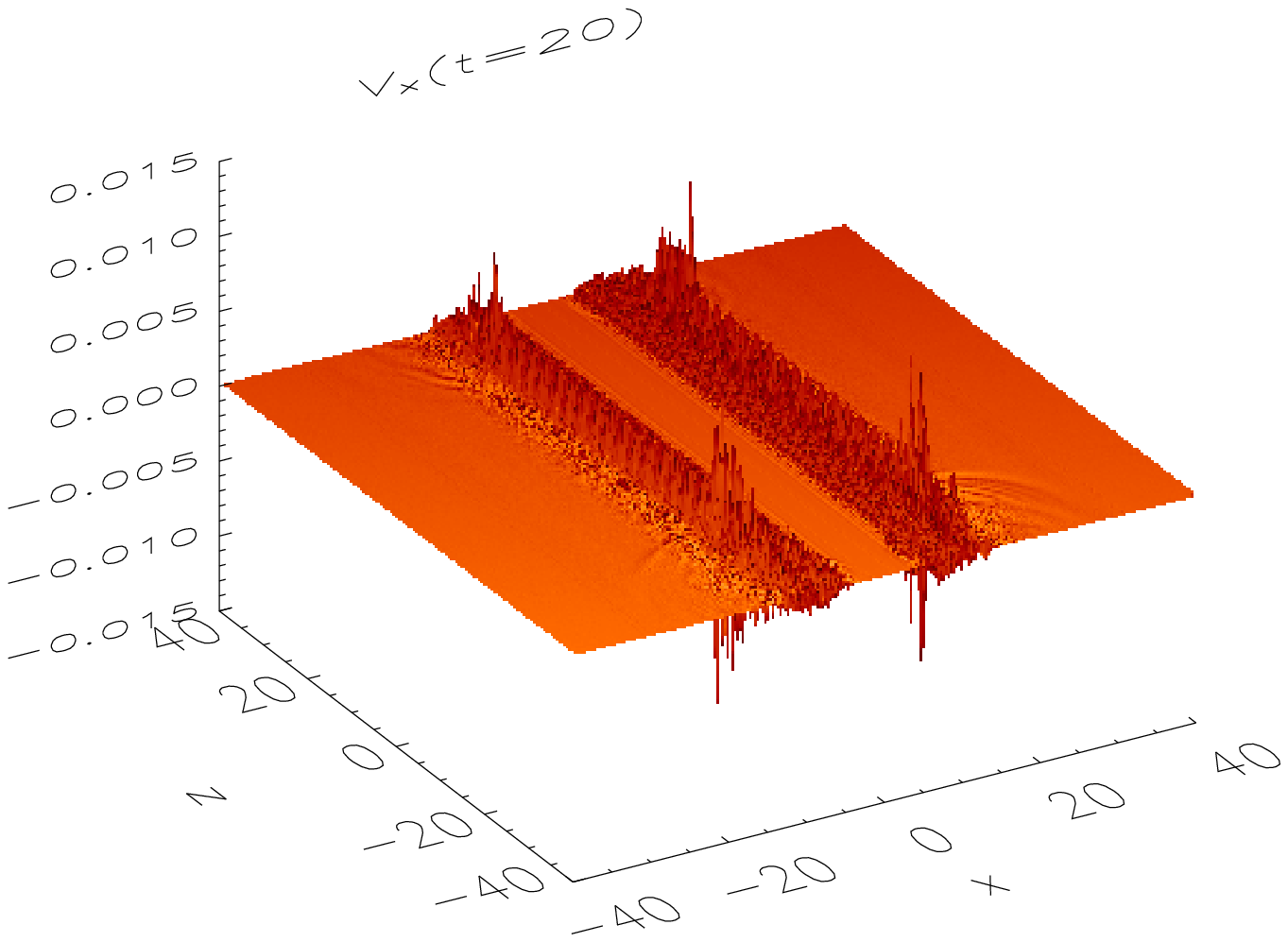,width=6.5cm}
 \epsfig{file=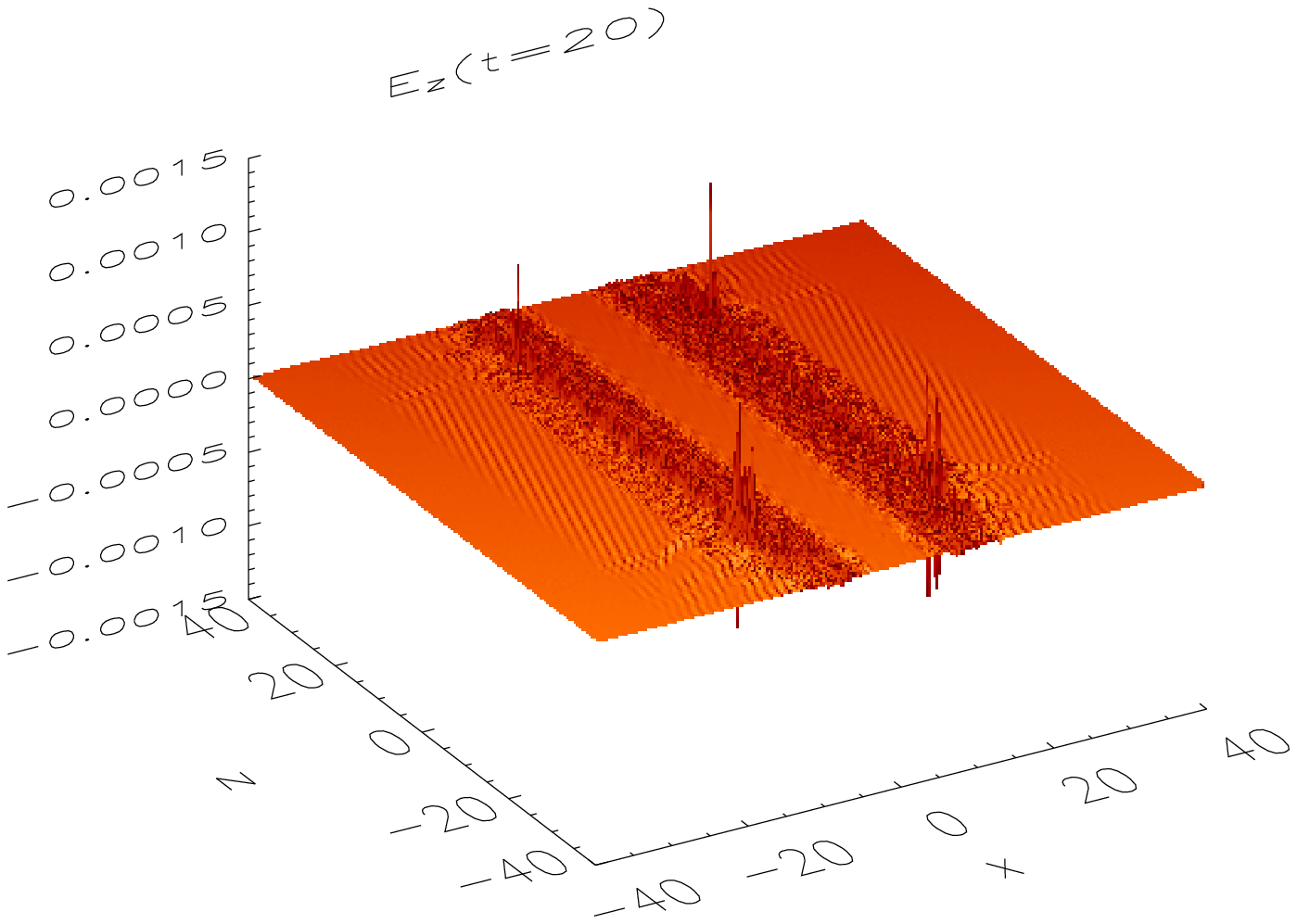,width=6.5cm}
 \caption{Snapshots of $V_x$ and $E_z$ at $t=2$ and 20 for the case of $k=10$, $\nu = 7$ Hz, $\lambda_A = 0.63$ Mm.}
\end{figure*}

\begin{figure}[]
\resizebox{\hsize}{!}{\includegraphics{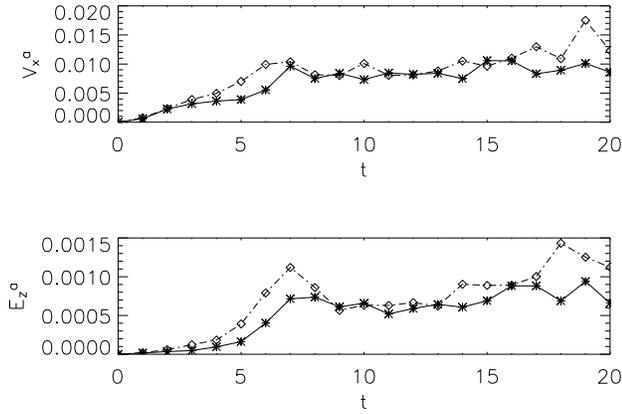}} 
\caption{Time evolution of the amplitudes of $V_x \equiv V_x^a$ and $E_z \equiv E_z^a$.
Solid lines with stars represent solutions using the Lare2d code with $4000 \times 4000$ resolution, while
dash-dotted lines with open symbols are the same but with $2000 \times 2000$ resolution. 
Here $k=10$, $\nu = 7$ Hz, $\lambda_A = 0.63$ Mm.}
\end{figure}

\begin{figure*}
\centering
\epsfig{file=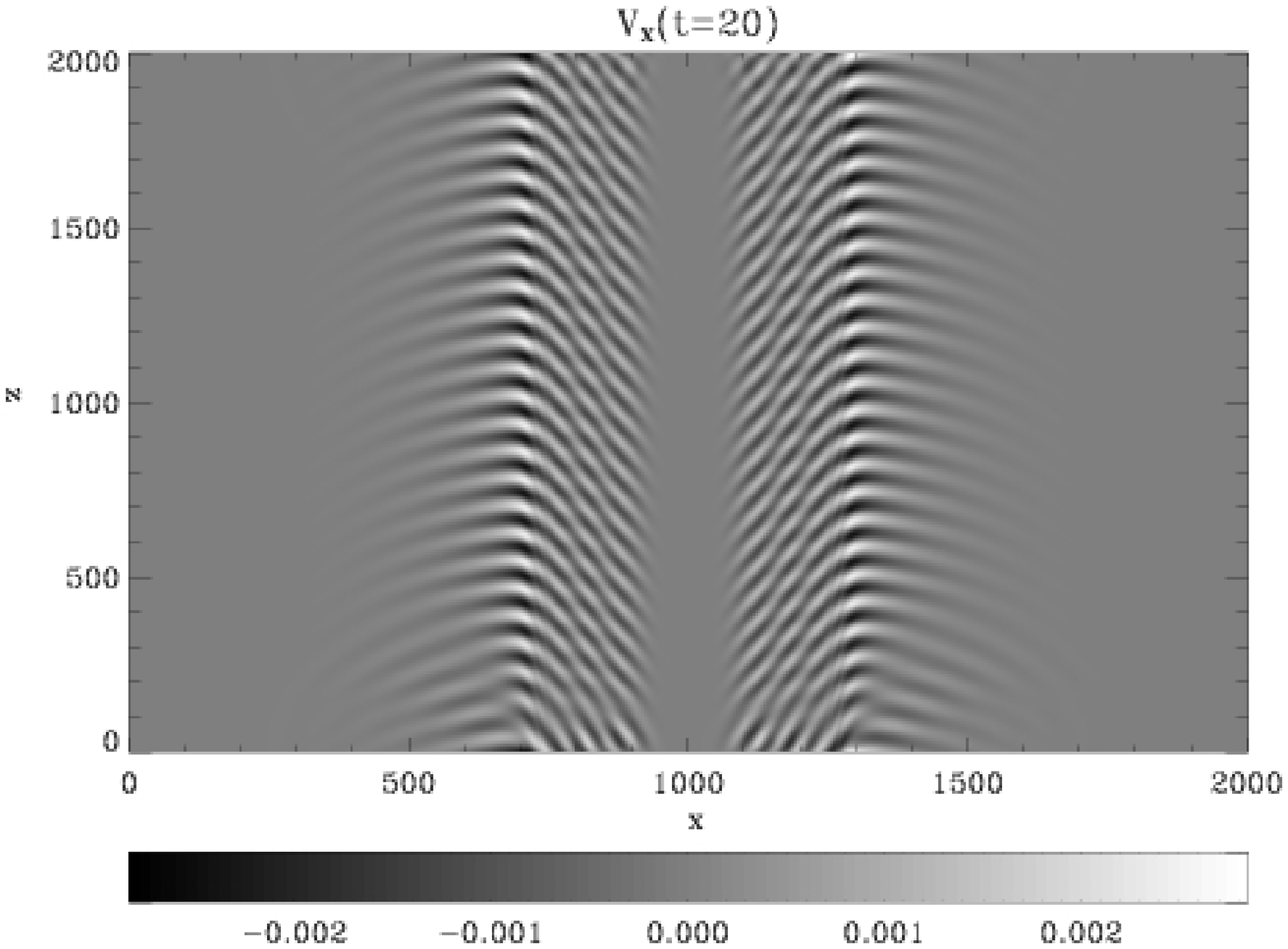,width=6.5cm}
 \epsfig{file=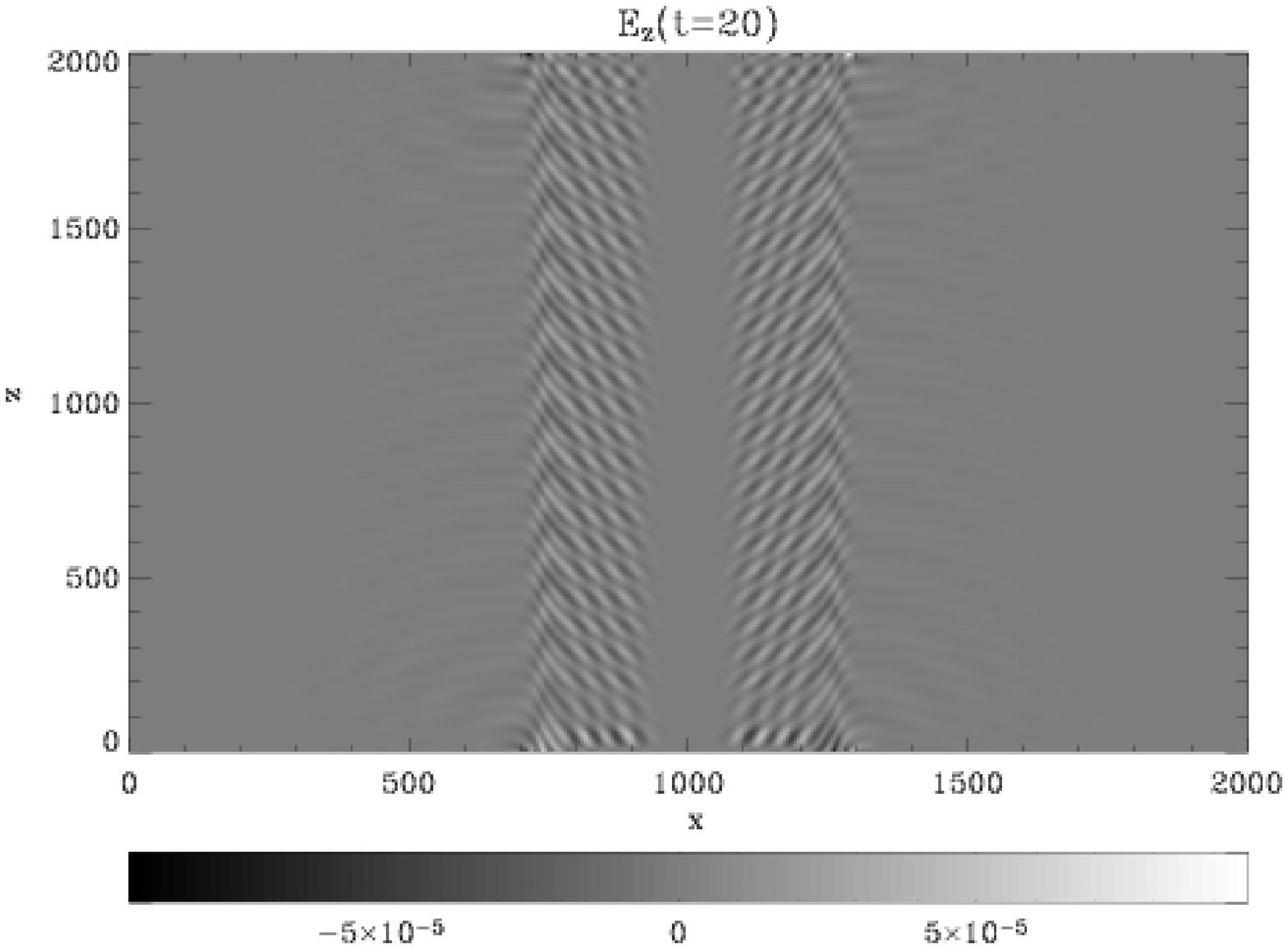,width=6.5cm}
  \caption{Intensity plots of $V_x$ and $E_z$ at $t=20$ for the case of $k=1$, $\nu = 0.7$ Hz, $\lambda_A = 6.3$ Mm.}
\end{figure*}

\end{document}